\documentstyle[preprint,aps]{revtex}
\preprint {IMSc-99/01/01}
\begin{document}
%@twocolumnfalse\endcsname \textwidth 16.5cm
%\textheight 21.0cm
\tightenlines
\draft
\title{Collective excitations in Quantum Dot}
\author{Subhasis Sinha}

\address
{The Institute of Mathematical Sciences, Madras 600 113, India.}
\date{\today}
\maketitle

\begin{abstract}

We investigate different types of collective excitations in a quantum dot
containing finite number of electrons at zero magnetic field. To estimate
the excitation energies analytically we follow the energy weighted
sum-rule approach. We consider the most general multipole excitation with
angular momentum $l$, and the breathing mode excitation (monopole
excitation) of a large quantum dot, for three different types of effective
electron-electron interaction. These are the logarithmic interaction, the
short range pseudopotential and the coulomb interaction. The ground state
density of the many-body system is calculated within Thomas-Fermi
approximation. The analytical results for the collective excitation energies
and their dependence on the system size and other external parameters are
discussed in detail. 
\end{abstract}


\section{Introduction}

With the advances in the nanofabrication technology, the study of the two
dimensional electron gas(2 DEG) has become an interesting topic. Quantum
dots are two-dimensional electrons confined within a finite area by
applying various gate voltages. It is an example of a finite size quantum
system where the number of particles can be varied from a few to few thousands. In
recent years, these systems have received considerable attention as a tool
to study the nature of electron-electron system in a finite size system.
In particular one of the goals has been to determine the far-infrared
response of the system and to understand the nature of collective
excitations. Though the collective excitations have been extensively
studied in other finite fermion systems like atomic
nuclei\cite{lipparini,stringari1} and metal clusters\cite{brack,deheer},
in three
dimensions, it still remains to be analysed in detail in
two-dimensional quantum dots. 

Recently, evidence has been found for a strong collective dipole mode and
its splitting in the presence of the magnetic field in far infrared
spectroscopy experiments\cite{merket,demel}. There is likely a further
evidence of a quadrupole excitation in a large quantum dot in the presence
of magnetic field\cite{demel} in experiments done using far infrared
spectroscopy. Recently collective spin and charge density excitations in a
quantum dot containing 200 electrons have been observed \cite{schuller}. 
Considerable amount of theoretical work has been
done to study the nature of edge manetoplasmon excitations of quantum dots in
strong magnetic fields\cite{tapash,serra,stringari2,mcdonald}.
The
collective excitations
have been studied within the
classical hydrodynamical model of quantum
dot\cite{fetter,sikin,peeters}. A lot of effort has been made to
understand the nature of the collective excitations in these
nanostructures by using different theoretical tools. 

In this paper we present detailed results on the nature of the collective
excitations in quantum dot systems at zero magnetic field. We consider the
multipole excitations and breathing modes of a quantum dot with finite
number of electrons, and calculate analytically the excitation energies
within sum rule approach. So far most of the microscopic calculations have
been done numerically, and the results are therefore restricted by the
particular choice of the device parameters. The main aim of this paper is
to investigate the collective excitations in a dot containing large number
of electrons analytically, to study the variation of the excitation
energies with the device parameters and with the change of electron
number. In a quasi zero dimensional nanostructure like quantum dot, the
nature of electron- electron in teraction is not very transperent. Other
than the conventional coulomb interaction, we consider two other types of
effective interactions. The advantage of obtaining an analytical
expression is the variation of the collective modes with the interaction
strength becomes clear. The ground state density profile
is evaluated using the simple Thomas-Fermi(TF) approximation. The T-F
approximation gives a good account of the bulk properties of the system
when the number of particles, $N$, is large\cite{matthias}. 
The T-F density deviates from the full
quantum mechanical density only in the region very close to the turning
point and it
neglects the asymptotic tail of the density. For the higher
multipolarities the density fluctuations shift more towards the edge. But
for
estimating the low lying multipole excitations of a large quantum system,
the T-F density is good enough and the edge correction of the density
gives rise to a small correction over the leading order estimate. We
estimate the edge correction for some important moments of density, and
the corrections typically goes as, $\sim c(l)/N^{3/4}$, where the term
$c(l)$ increases with increasing multipolarity $l$. This estimate of the
edge correction justifies the evaluation of different moments of a large
quantum system by using T-F density. So that the low-lying collective
excitations of a large dot can be estimated within the T-F approximation.
 
For the
electron-electron interaction we have taken three different forms of the
two-body interaction. These forms are chosen so as to facilitate
analytical calculations of excitation energies. However, we also point out
that in some limits these approximate the effective electron-electron
interaction in a quantum dot. Since the effective interaction changes with the 
number of particles, it is useful to work with different approximate 
interactions. 
  
This paper is organized as follows: In section II we discuss the ground
state of the quantum dot in the simple TF approximation assuming various
forms of the electron-electron interaction. In Sec.III we discuss the RPA
dynamic polarisibility of quantum dot for different excitation operators,
and the corresponding strengh distributions of such excitations. Also we
discuss how the collective excitation energy can be estimated using the
moments of the corresponding strength distribution. We explicitely
evaluate the multipole excitation energy for different models of quantum
dot in Sec.IV. In Sec.V we derive the sum rules and excitation energy for
the breathing mode of the dot, from simple scaling argument. Section VI
contains a summary. In Appendix A and Appendix B, we outline the 
corrections that may arise from using T-F density for the ground state. 

\section{Thomas-Fermi method for ground state}

The two dimensional Hamiltonian to describe the many electrons in a 
quantum dot may be written as
\begin{equation}
H = \sum_{i=1}^{N} [\frac{p_{i}^{2}}{2m^{*}} + \frac{1}{2}m^{*}\omega^{2} 
r_{i}^{2} + \sum_{j>i=1}^{N} V(|\vec{r_{i}} - \vec{r_{j}}|)].
\end{equation}
Here we have assumed that the electrons are confined in a parabolic 
potential which is a good approximation to the device potential, $m^{*}$ 
is the effective mass of the electron in GaAs sample, which is $0.067$ 
times the actual electron mass. We estimate the density profile of the 
many-body system by using T-F energy functional. It is well known that for 
the large number particles, T-F density profile agrees very well with the 
full quantum mechanical density, except very close to the turning 
point. Asymptotic tail of the density profile can be obtained by including 
the gradient corrections in the density functional, which includes the higher 
order $\hbar$ corrections in the T-F energy functional. In evaluating the 
energy weighted sum rules, T-F density gives the leading order contribution, 
and the small edge correction can give rise to the higher order $(1/N)$ 
corrections. In what follows, we assume that $N$ is sufficiently large so
that these corrections may be neglected.
The energy functional describing the ground state of this many body system is
\begin{equation}
E[\rho] = \int d^{2}r [\tau(r) + V_{sc}(r)\rho(r)],
\end{equation}
where the self-consistent potential $V_{sc}$ is given by,
\begin{equation}
V_{sc} = \frac{1}{2}m^{*}\omega^{2} r^{2} + \frac{1}{2}\int d^{2}r 
V(|\vec{r} - \vec{r'}|) \rho(r').
\end{equation}
Within  the Thomas-Fermi approximation the kinetic-energy density is given by,
$$\tau(r) = \frac{\hbar^{2} \pi}{2 m^*} \rho^{2}(r),$$
for spin $1/2$ 
electrons in an unpolarised dot. The self-consistent equation for 
determining the density is obtained by minimising the free energy $$F = (E - 
\mu N),$$ where $\mu$ is the chemical potential. Thus we have the 
following self-consistent equation for the density,
\begin{equation}
\frac{\hbar^{2} \pi \rho(r)}{m^*} + \frac{1}{2}m^{*} \omega^{2} r^{2} + 
\int d^{2}r V(|\vec{r} - \vec{r`}|) \rho(r') = \mu.
\end{equation}
The chemical potential is determined by conserving the number of 
particles in the dot,
\begin{equation}
\int d^{2}r \rho(r) = N.
\end{equation}

In the following subsections we consider three different forms of the
repulsive two body interaction and try to find out self-consistent density
analytically. We first show that for logarithmic interaction and
short-range interaction, the self-consistent density can be solved
exactly. However, for coulomb interaction the above integral equation can
not be solved exactly.  Therefore we estimate the ground-state
density profile by variational method. 

\subsection{Logarithmic interaction}

The two dimensional Hamiltonian with logarithmic interaction is given by,
\begin{equation}
H = \sum_{i=1}^{N} [ \frac{p_{i}^{2}}{2 m^*} + \frac{1}{2} m^* \omega^{2}
r_{i}^{2} ] - g \sum_{i<j}^{N} log[\frac{|\vec{r_{i}} - \vec{r_{j}}|}{a}].
\end{equation}
Here, $a$ is an arbitary parameter which determines the range of the
two-body interaction, and $g$ is the strength of the interaction. The
advantange of taking logarithmic two-body interaction is that the
Thomas-Fermi model is exactly solvable \cite{pino,sinha}, and it can
mimic the realistic two body interaction. This model may also be justified
from the finite thickness effect of the two dimensional system
\cite{sinha}. As shown by Zhang and DasSarma\cite{zhang}, the effective
interaction for a quasi two-dimensional electron system  in momentum 
space is given by, 
\begin{equation}
V(k) = \frac{e^{2}}{2 \pi \epsilon k} \frac{1 +9 k/8b_{z} + 3
k^{2}/8b_{z}^{2}}{(1 + k/b_{z})^{3}}
\end{equation}
where, $b_{z}$ has dimensions of inverse length. The effective potential
given above behaves approximately as $log(r)$ over small distances and
$\frac{1}{r}$ for large distances. The strength of the logarithmic
interaction can be written as $g \approx \frac{3 e^{2} b_{z}}{8\epsilon}$. 

Following Thomas-Fermi method the density distribution of the electron
system can be determined from the following equation,
\begin{equation}
\frac{\pi \hbar^{2} \rho(r)}{m^{*}} + \frac{1}{2} m^* \omega^{2} r^{2} - g
\int d^{2}r' \rho(r') ln|\frac{\vec{r} - \vec{r'}}{a}| = \mu.
\end{equation}
Now applying laplacian operator on both sides of the equation we obtain
the following differential equation,
\begin{equation}
\frac{\hbar^{2} \pi}{m^*} \nabla^{2} \rho(r) + 2 m^{*} \omega^{2} - 2 \pi
g \rho(r) = 0.
\end{equation}
Assuming the circular symmetry for the density of the ground state, the 
above equation can be writen as,
\begin{equation}
\frac{d^{2} y}{dx^{2}} + \frac{1}{x} \frac{dy}{dx} - 2 \tilde{g} y = 0,
\end{equation}
where, the dimensionless variable $ x = r/l_{0}$ and
$y = \rho(r) -\frac{1}{\pi l_{0}^{2} \tilde{g}}.$ 
The harmonic 
oscillator length scale and the
dimensionless coupling are defined as, $l_{0} = \sqrt{\frac{\hbar}{m^{*}
\omega}}$ and $\tilde{g} = \frac{g}{\hbar \omega}$.
The solution of the above differential equation is given in terms of 
modified Bessel functions, 
\begin{equation}
y(x) = A I_{0}(\sqrt{2 \tilde{g}} x) + B K_{0}(\sqrt{2 \tilde{g}} x),
\end{equation}
where $A$ and $B$ are constants to be fixed by the boundary conditions. In
the limit $x \rightarrow 0$, $I_{0} \rightarrow 1$,  where as $K_{0}
\rightarrow -ln(x)$ \cite{grad}. The latter condition implies that the
density at the
centre is infinite which is unphysical since it would require infinite
energy to push particles to the centre due to the Coulomb repulsion.
Obviously minimisation of the ground state energy implies that $B$ should
be zero. The quantity $A$ is then fixed by the boundary condition,
$\rho(r_{0}) = 0$ at the turning point. We have therefore the full
solution for the TF density,
\begin{equation}
\rho(r) = \frac{1}{\pi l^{2} \tilde{g}} [ 1 - \frac{I_{0}(\sqrt{2
\tilde{g}} x)}{I_{0}(\sqrt{2 \tilde{g}} x_{0})}],~~~~~~~~for~ x \leq
x_{0}.
\end{equation} 
The turning point $r_{0}$ can be determined from the normalisation
condition,
\begin{eqnarray}
N & = & \int \rho(r) d^{2}r \nonumber \\
  & = & \frac{1}{\tilde{g}^{2}} [ \frac{\tilde{L}^{2}}{2} - \tilde{L}
\frac{I_{1}(\tilde{L})}{I_{0}(\tilde{L})} ]
\end{eqnarray}
here $ \tilde{L} = \sqrt{2 \tilde{g}} x_{0}$, and $x = r/r_{0}$.
For $\tilde{L} >> 1$,(which is equivalent to large N limit) the density 
becomes almost constant at the value $\rho_{0} = \frac{1}{\pi l_{0}^{2} 
\tilde{g}}$, and the relation between $\tilde{L}$ and the number of 
particles N becomes,
\begin{equation}
2 \tilde{g} N = \tilde{L}^{2} - 2 \tilde{L}.
\end{equation}
In the other limit $\tilde{L} << 1$ (which is equivalent to the weak 
coupling limit), the density becomes almost parabolic, and the dependence 
of $\tilde{L}$ on N is given by,
\begin{eqnarray}
\tilde{g}^{2} N & = & \frac{\tilde{L}^{4}}{16}(1 +
\frac{1}{12}\tilde{L}^{2})\nonumber\\
N & = & \frac{1}{4}x_{0}^{4}(1 + \frac{1}{6}\tilde{g} x_{0}^{2}).
\end{eqnarray}
In fig.1 we have compared the scaled density profile of the electrons with
logarithmic interaction with the parabolic one, for two different values
of the parameter $\tilde{L}$. When $\tilde{L} = 1$, then the density
becomes almost parabolic, but for $\tilde{L} = 20$, the density becomes
very flat.

 \subsection{Short range interaction}

We now consider the case when electrons in the dot are interacting by a very short 
range delta function interaction,$V(r) = V_{0} \delta(r)$.
For a dilute Fermi gas the two body interactions can be approximated by
delta function pseudopotential. The interaction strength $V_{0}$ is related
to the s-wave scattering length $a$ in the following way, $V_{0} =
\frac{4 \pi \hbar^{2} a}{m^{*}}$.
Also it has been found that for 2-D electron gas in strong magnetic field
the short range pseudo-potential is a very good approximation \cite{trugman}.
This system is also interesting because short range interaction among the 
fermions can mimic exclusion statistics\cite{murthy}.

The energy functional of the electrons with short range two body
interaction can be written as,
\begin{equation}
E[\rho] = \int d^{2}r [\frac{\hbar^{2} \pi}{2 m^{*}} \rho^2(r) +
\frac{1}{2} m^{*} \omega^{2} r^{2} \rho(r) + \frac{V_{0}}{2} \rho^2(r)].
\end{equation}
By minimizing the energy functional with respect to the density, we obtain
the following equation,
\begin{equation}
\frac{\hbar^{2} \pi}{m^{*}} \rho(r) + \frac{1}{2} m^{*} \omega^{2} r^{2} +
V_{0} \rho(r) = \mu,
\end{equation}
where $\mu$ is the chemical potential of the system. The self consistent
density can be written as,
\begin{equation}
\rho(r) = \frac{1}{\pi l_{0}^{2} \alpha_{s}}(\tilde{\mu} - \frac{1}{2}
x^{2}),
\end{equation}
where, $x$ is the dimensionless variable $x = r/l_{0}$ and other parameters
are
defined as, $\alpha_{s} = 1 + \frac{V_{0} m^{*}}{\pi \hbar^{2}}$ and
$\tilde{\mu} = \frac{\mu}{\hbar \omega}$.
The turning point of the density distribution can be determined from the
conservation of the total number of particles,
\begin{equation}
N = \int d^{2}r \rho(r) =  \frac{x_{0}^{4}}{4 \alpha_{s}}.
\end{equation}

The results thus obtained with short range interaction are similar 
to the non-interacting but confined electrons. The only difference with the 
non-interacting electrons is the scaling factor $1/\alpha_{s}$ both in 
density and in the number of particles.

\subsection{Coulomb interaction}
Next we consider the case where the two-body electron electron
interaction is $1/r$. For this form of the two-body interaction, T-F self
consistency equation for density can not be solved analytically. To do the
rest of the calculations analytically, we take a variational ansatz for
the self consistent density profile and determine the parameters of the
ansatz by minimising the energy functional. In zero magnetic field we 
can take the following ansatz for the density profile,
\begin{equation}
\rho(r) = \frac{1}{2 \pi l_{0}^{2} \alpha_{c}} (x_{0}^{2} - x^{2}),
\end{equation}
where, $\alpha_{c}$ is the variational parameter, and $x$ is the 
dimensionless
variable $r/l_{0}$. The density of the non interacting electron gas 
corresponds to the limit $\alpha_{c} = 1$. The inclusion of the two body 
repulsion flattens the density profile, which can be thought of as the 
reduction of the frequency $\omega_{eff}$ of the effective mean field 
potential. So the electron-electron interaction makes $\alpha_{c} > 1$. In 
the case of short-range interaction the density profile is parabolic, and 
also for the logarithmic interaction density profile becomes parabolic in the 
strong confinement limit. We expect at zero magnetic field the ansatz can 
give good estimate of the ground state density in the strong confinement 
limit. We must, however, caution that this ansatz breaks down in the 
presence of 
a strong magnetic field. The ground state density here deviates from
the parabolic form and is approximately $\rho(r) \sim \sqrt{1 -
(r/R)^2}$ \cite{fogler}. For convenience we define $ \rho_{0} = \frac{1}{2
\pi l_{0}^{2} \alpha_{c}}$. The turning point $x_{0}$
can be
calculated from the total number of electrons $N$,
\begin{equation}
N  =  \int d^{2} r \rho(r) ~~~~~~
 =  \frac{1}{2} \pi l_{0}^{2} \rho_{0} x_{0}^{4}.
\end{equation}
Let us calculate the contributions to the energy coming from the different 
parts of the density functional as a function of the parameter $\rho$.
Contribution to the kinetic energy is given by,
\begin{equation}
E_{\tau} = \frac{\hbar^{2} \pi}{2 m^{*}} \int d^{2}r \rho^2(r)=
\frac{\hbar^{2} \pi^{2} \rho_{0}^{2} l_{0}^{2} x_{0}^{6}}{6 m^{*}}.
\end{equation}
The potential energy due to the external confinement is given by,
\begin{equation}
E_{p} = \frac{1}{2} m^{*} \omega_{0}^{2} \int d^{2} r r^{2} \rho(r)
= \frac{1}{12} \pi m^{*} \omega_{0}^{2} \rho_{0} l_{0}^{4} x_{0}^{6}.
\end{equation}
The interaction energy within the Hartree approximation is given by,
\begin{equation}
E_{H} = \frac{1}{2} \int d^{2}r d^{2} r' \rho(r)V(|\vec{r} -
\vec{r'}|)\rho(r),
\end{equation}
where, $V(|\vec{r} - \vec{r'}|) = \frac{e^{2}}{\epsilon} \frac{1}{|\vec{r}
- \vec{r'}|}$. In
momentum space the above expression can be written as,
\begin{equation}
E_{H} = \frac{1}{2} \frac{e^{2}}{2 \pi \epsilon} \int d^{2}q \frac{1}{q} 
\rho(q) \rho(-q),
\end{equation}
where,
\begin{equation}
\rho(q) = \int \rho(r) e^{i\vec{k}.\vec{r}} d^{2} r = 4 \pi \rho_{0} 
x_{0}^{2} \frac{J_{2}(q r_{0})}{q^{2}}.
\end{equation}
After performing  the integration we obtain the Hartree energy in the 
following form,
\begin{equation}
E_{H} = \frac{\Gamma(\frac{1}{2}) \Gamma(4) r_{0}^{3}}{2 \Gamma(5/2) 
\Gamma(9/2) 
\Gamma(5/2)} \frac{e^{2}}{\epsilon} (\pi x_{0}^{2} \rho_{0})^2.
\end{equation}
In terms of the variational parameter $\alpha_{c}$ and the number of 
particles $N$, the total energy of the system is given by,
\begin{equation}
E_{0} = \frac{1}{3} \hbar \omega_{0} N^{3/2} (\alpha_{c}^{1/2} + 
\frac{1}{\alpha_{c}^{1/2}}) + \frac{512}{315} \frac{\sqrt{2}}{\pi 
\alpha_{c}^{1/4}} \frac{e^{2}}{\epsilon l_{0}} N^{7/4}.
\end{equation}
To get the minimum energy, the total energy is be minimized with respect 
to the variational parameter $\alpha_{c}$ when the number of particles $N$
is 
held fixed. The $\alpha_{c}$ that one obtains at the minimum of the energy
is 
therefore a function of $N$, among other things. This is in contrast to the 
short-range interaction where $\alpha_{s}$ is independent of $N$.

Having thus determined the ground state energy and the density in the TF 
approximation, we now use these in evaluating the collective excitation 
energies for different types of collective excitations.

\section{Sum rules and collective excitations}

The excitation spectrum of any system is usually probed by applying
external fields. The effect of the perturbing probe can be described
through an interaction Hamiltonian,
\begin{equation}
H_{int} = \lambda (F e^{i \omega t} + F^{\dagger} e^{-i \omega t}),
\end{equation}
where the quantity $\lambda$ gives the strength of the oscillating field. 
Given an excitation operator $F$, many useful quantities of the perturbed
system can be calculated from so-called strength function,
\begin{equation}
S_{\pm} = \sum_{n} |<n| F_{\pm} |0>|^{2} \delta(E - E_{n}),
\end{equation} 
where, $E_{n}$ and $|n>$ are the excitation energy and excited state
respectively, and $F_{+} = F$, $F_{-} = F^{\dagger}$. Various sum rules 
are derived through the moments of the strength function, defined as,
\begin{eqnarray}
m^{\pm}_{k} & = & \frac{1}{2} \int E^{k} (S_{+}(E) \pm S_{-}(E)) dE
\nonumber \\
& = & \frac{1}{2} ( <0|F( \hat{H} - E_{0})^{k} F^{\dagger} |0> \pm
<0|F^{\dagger} (\hat{H} - E_{0})^{k}F |0> ). 
\end{eqnarray}
It is easy to see that, for a given $k$, the moments may be expressed in 
terms of the commutators of the excitation operator $F$ with the many 
body Hamiltonian $H$. We give below some of the useful sum rules,
\begin{eqnarray} 
m^{-}_{0} & = & \frac{1}{2} <0|[F^{\dagger}, F]|0>, \\
m^{+}_{1} & = & \frac{1}{2} <0|[F^{\dagger}, [H, F]]|0> \\
m^{-}_{2} & = & \frac{1}{2} <0|[J^{\dagger}, J]|0> \\
m^{+}_{3} & = & \frac{1}{2} <0|[J^{\dagger}, [H, J]|0>~;~~~~J = [H, F].
\end{eqnarray}

The choice of the excitation operator $F$ is dictated by the physics that
one wants to describe.  In this paper we consider two types of collective
excitations:  the multipole excitation modes and the breathing modes in a
the quantum dot. In two-dimensions multipole excitation operator can be
written as $F = \sum_{i}r_{i}^{l} e^{i l \theta_{i}} = \sum_{i}z_{i}^l$,
where $z$ is the complex
variable $x + i y$. Angular part of any function in two dimensions can be
expanded in terms of these operators. These multipole modes describe the
angular momentum excitations of the system. The excitation operator for
breathing mode is given by $ F = \sum_{i} r_{i}^{2}$. The collective
excitation energy can be estimated from the sum-rules of the corresponding
excitation operators. 

For a highly collective state the strength distribution becomes a sharply
peaked function around the collective excitation energy. If the excitation
operator $F$ is hermitian, or if the Hamiltonian does not contain any time-reversal
symmetry breaking term (as in the quantum dot without magnetic field), then we can 
take $m_{2}^{-} = 0$ and the low lying collective excitation energy is given by, 
\begin{equation}
E_{c} =
\sqrt{\frac{m^{+}_{3}}{m^{+}_{1}}}. 
\end{equation}
Near the collective excitation energy, we can approximate the strength
distribution by a delta function $S_{\pm} (E) = \sigma_{\pm} \delta(E -
E_{c})$, then we can trivially show the above form of collective
excitation. But in general the strength function has a finite width.
Following ref.\cite{serra}, one can then derive the above form of 
collective 
excitation energy for multipole excitations by using the variational 
principle. 
Given the $N$ electron ground state $|0>$, it is possible to find the
collective excitation energy and the collective state $|c>$, if one is
able to find an operator $O^{\dagger}$, which satisfies the following
equation of motion, 
\begin{equation} [\hat{H}, O^{\dagger}] = \hbar
\omega_{coll} O^{\dagger}. 
\end{equation} 
The state $O^{\dagger}|0>$ has
excitation energy $\hbar \omega_{coll}$.  The excitation energy is then 
given by the following expression,
\begin{equation} 
\hbar \omega_{coll} =
\frac{<0|[O, [\hat{H}, O^{\dagger}]]|0>}{<0|[O, O^{\dagger}]|0>}.
\end{equation} 
We may now take the variational ansatz for $O^{\dagger}$
as, $ O^{\dagger} = F + aJ$  with $a$ as the variational parameter. 
Substituting this expression in the above equation, we obtain the
excitation energy as, 
\begin{equation} \hbar \omega_{coll} =
\frac{m^{+}_{1} + a^{2} m^{+}_{3}}{2 a m^{+}_{1}}, 
\end{equation}
where we
have assumed $m^{-}_{2} = 0$, for the system in the absence of the  
magnetic field. By
minimizing the energy with respect to the variational parmeter $a$, we
obtain the collective excitation energy as 
$\hbar \omega_{coll} =\sqrt{\frac{m^{+}_{3}}{m^{+}_{1}}} = E_{c}$.

\section{Multipole excitations in Quantum dot}

The multipole excitations in a quantum dot lead to the change in total
angular momentum.  In particular, the 
dipole excitation in parabolic
quantum dot has been well studied both theoretically and experimentally.
The corresponding operator for the dipole excitation operator is given by 
$F= x + iy$. One can generalise this and write for any multipole excitation
operator $F_{l} = z^{l}$. We use this form to calculate the relevant 
sum-rules for multipole excitations.
The first moment of the strength distribution function may be written as,
\begin{eqnarray}
m^{+}_{1}(l) & = & \frac{1}{2}<0| [\sum z_{i}^{l}, [\hat{H}, \sum 
z_{i}^{l}]]|0> \nonumber \\ & = & \frac{\hbar^{2} l^{2}}{m^{*}} \int
d^{2}r r^{2l - 2} \rho(r),
\end{eqnarray}
where $\rho(r)$ is the ground state density distribution.  

The calculation of $m_3^+(l)$ is more complicated. To make the analysis 
transparent, we consider the various contributions to the third moment 
separately. The sum-rule 
$m^{+}_{3}(l)$ may then be written in the following form, 
\begin{equation}
m^{+}_{3}(l) = m_{3}(T) + m_{3}(V) + m_{3}(ee),
\end{equation}
where the first term denotes the contribution from the kinetic energy 
operator in $H$, the second term denotes the contribution from the 
confinement potential, and the last term is the contribution from the 
interaction part. We now calculate separately each one of these 
contributions.

The kinetic energy term in $m_{3}(T)$ is given by,
\begin{equation}
m_{3}(T) = 2 \hbar^{2} (l -1) \frac{\hbar^{2} l^{2}}{m^{*2}} \int
d^{2}r [ r^{2(l - 2)} (l \tau(r) + 2 (l - 2) \lambda(r)],
\end{equation}
where, $\tau(r) = <\frac{\vec{p}^{2}}{2 m^{*}}>$, and $\lambda(r) = <
\sum_{i} \frac{\hbar^{2} l'^{2}}{2 m^{*} r^{2}}> = < \frac{(\vec{r} \times
\vec{P})^{2}}{2 m^{*} r^{2}}>$. Here the angular brackets denotes the
averages over the momentum distribution. Note that unlike the ground 
state where only the first term contributed, for the multipole 
excitations the centrifugal barrier, denoted by the second term, 
contributes to the sum rules.
Using the Thomas-Fermi approximation this contribution is givenby, 
$\lambda(r) =
\frac{1}{2} \tau(r) = \frac{\hbar^{2}\pi}{4 m^{*}} \rho^2(r)$.
Substituting these expressions in the above equation we obtain,
\begin{equation}
m_{3}(T) = \frac{2 \hbar^{4} \pi (l - 1)^{2}}{m^{*}} \frac{\hbar^{2}
l^{2}}{m^{*2}} \int d^{2}r r^{2(l - 2)} \rho^2(r).
\end{equation}
The contribution coming from the harmonic oscillator cofinement in the
sum-rule is given by,
\begin{equation}
m_{3}(V) = l m^{*} \omega_{0}^{2} \hbar^{2} \frac{\hbar^{2}
l^{2}}{m^{*2}} \int d^{2}r r^{2 (l - 1)} \rho(r).
\end{equation}   
The last term $m_{3}(ee)$ due to two body interaction can be written as,
\begin{eqnarray}
m_{3}(ee) & = & \frac{1}{2} \hbar^{2} \frac{\hbar^{2} l^{2}}{m^{*2}} [\int
d^{2}r
V_{H}(r) r^{2(l - 1)} \rho^{''}(r) + (2l - 1) \int d^{2}r V_{H}(r) r^{2l -
3} \rho'(r) \nonumber \\ & &  + \int d^{2}r \int d^{2}r' \rho'(r) r^{l -
1}
e^{-i l \theta}
V(|\vec{r} - \vec{r'}|) e^{i l \theta'} r'^{l - 1} \rho'(r')],
\label{m3int}
\end{eqnarray}
where $V(r)$ is the two body interaction, $V_{H} = \int d^{2}r V(|\vec{r}
- \vec{r'}|) \rho(r')$, and $\rho'(r) = \frac{\partial \rho(r)}{\partial
r}$.

We thus have the general expressions for the sum rules where the only 
unknown is the ground state density $\rho(r)$. For specific interactions, 
the ground state density was calculated in Section II. We now use this to 
explicitly compute the sum rules. Note however, that the T-F density
calculated in section II fails near the turning point, since the T-F
density abruptly goes to zero. But for the lower multipolarities main
contribution to the moments comes from the bulk region, and the diffusive
edge rigion gives rise to small correcction, which decreases with the
increasing $N$. We discuss the corrections to this approximation in the
appendices.

\subsection{Logarithmic potential}

For logarithmic potential the first moment $m^{+}_{1}$ is given by,
\begin{eqnarray}
m^{+}_{1} & = & m^{*} \frac{\hbar^{2} l^{2}}{m^{*2}} \int d^{2}rr^{2(l 
- 1)} \rho(r) \nonumber \\
 & = & m^{*}(\frac{\hbar^{2} l^{2}}{m^{*2}}) \rho_{c} r_{0}^{l} 2\pi
\int_{0}^{1} dx x^{2l - 1} [1 -
\frac{I_{0}(x \tilde{L})}{I_{0}(\tilde{L})}],
\end{eqnarray}
where the central density $\rho_{c} = \frac{1}{\pi l_{0}^{2} \tilde{g}}$
and  
the dimensionless variable $x = r/r_{0}$, and $\tilde{L} = \sqrt{2
\tilde{g}} x_{0}$. 

In the case of the third moment $m^{+}_{3}$, kinetic energy and 
the confinement energy contributions can be written as,
\begin{equation}
m_{3}(T) + m_{3}(V) = \frac{\hbar^{2} l^{2}}{m^{*2}} [ 2 (l - 1)^{2} 
\frac{\hbar^{4} \pi}{m^{*}} \int d^{2}r r^{2(l - 2)} \rho^{2}(r) + l 
m^{*} \omega^{2} \hbar^{2} \int d^{2}r r^{2(l - 1)} \rho(r)].
\end{equation}
We now evaluate the remaining term $m_{3}(ee)$. Integrating 
eq.(\ref{m3int}) by parts we can write $m_{3}(ee)$ as,
\begin{equation}
m_{3}(ee) = \frac{\hbar^{4} l^{2}}{2 m^{*2}}[ \int d^{2}r
\nabla^{2}V_{H} 
r^{2(l - 1)} \rho(r) + 2(l - 1)\int d^{2}r \frac{\partial V_{H}}{\partial 
r}r^{(2l -3)} \rho(r) + I_{3}],
\end{equation}
where, $V_{H}(r)$ is the hartree potential. For logarithmic 
interaction we can use the following relations,
\begin{eqnarray} 
\nabla^{2} V_{H}(r) =
-2\pi g \rho(r), \\ 
\frac{\partial V_{H}}{\partial r} = - m^{*}
\omega^{2} r - \frac{\hbar^{2} \pi}{m^{*}} \frac{\partial \rho}{\partial r}. 
\end{eqnarray}
Therefore the term $I_{3}$ is given by,
\begin{eqnarray}
I_{3} & = & \int d^{2}r \int d^{2}r \rho(r)'~r^{l - 1} e^{-i l \theta}
V(|\vec{r} - \vec{r'}|) e^{i l \theta'} r'^{l - 1} \rho'(r') \nonumber \\
& = & (2 \pi)^{2} \int d^{2}q f(q) \int dr r^{l} J_{l}(q r) \rho'(r) \int
dr' r'^{l} J_{l}(q r') \rho'(r'),
\end{eqnarray}
where $f(q)$ is the two body potential in momentum space, and $\rho'(r) =
\frac{\partial \rho(r)}{\partial r}$. After doing the integration by parts
and using some identities for Bessel functions we can rewrite the above
expression in the following form,
\begin{eqnarray}
I_{3} & = & \int d^{2}q q^{2} f(q) \int d^{2}r e^{i \vec{q}.\vec{r}}
e^{-i(l -
1) \theta} r^{(l - 1)} \rho(r) \int d^{2}r' e^{-i\vec{q}.\vec{r'}}
e^{i
(l - 1)
\theta'} r'^{(l - 1)}\rho(r') \nonumber \\
& = & 2 \pi g \int d^{2}r r^{2 (l - 1)} \rho^{2}(r). 
\end{eqnarray}
Using the above results we can write the multipole excitation energy in
the following form,
\begin{equation}
E_{c}^{2} = \frac{m_{3}^{+}}{m_{1}^{+}} = \hbar^{2} \omega^{2} +
\frac{3 \hbar^{4} \pi (l - 1)^{2}}{m^{* 2}} \frac{\int d^{2}r r^{2(l -
2)}
\rho^{2}(r)}{\int d^{2}r r^{2(l - 1)} \rho(r)}.
\end{equation}
The above equation can be written as,
\begin{equation}
E_{c}^{2} = \hbar^{2}\omega^{2}[1 + 6 l (l - 1) 
\frac{f(\tilde{L})}{\tilde{L}^{2}}],
\end{equation}
where,
\begin{equation}
f(\tilde{L}) = (\frac{l - 1}{l}) \frac{\int_{0}^{1}dx x^{2l - 3}[1 - 
\frac{I_{0}(x\tilde{L})}{I_{0}(\tilde{L})}]^{2}}{\int_{0}^{1}dx x^{2l - 1}
[1 - \frac{I_{0}(x \tilde{L})}{I_{0}(\tilde{L})}]}.
\end{equation}
In the large $\tilde{L}$ limit, one can approximate the function $[1 - 
\frac{I_{0}(x\tilde{L})}{I_{0}(\tilde{L})}]$ by a simple form $[1 - 
exp(\tilde{L}(x - 1))]$, and the approximate value of $f(\tilde{L})$ is given by,
\begin{equation}
f(\tilde{L}) = \frac{1 - 2 F_{1}(2l - 2, 2l - 1; \tilde{L})e^{-\tilde{L}}
+ F_{1}(2l - 2, 2l - 1;2\tilde{L})e^{-2\tilde{L}}}{1 - F_{1}(2l, 2l +
1;\tilde{L})e^{-\tilde{L}}},
\end{equation}
where $F_{1}$ is the degenerate hypergeometric series \cite{grad}. In the
large
$\tilde{L}$ limit asymptotic $N$ dependence of the collective excitations
are given by,
\begin{equation}
E_{c} = \hbar \omega \sqrt{1 + \frac{3l(l - 1)}{\tilde{g}^{2} N}}.
\end{equation}
 By substituting $l = 1$, for dipole excitation we can see that the
excitation energy is $\hbar \omega$, which obeys Kohn's theorem
\cite{kohn}. In the other limit $\tilde{L} << 1$, (which is the weak coupling
limit) the dispersion relation of the multipole modes are given by,
\begin{equation}
E_{c} = \hbar \omega [3l -2 - 6 \frac{(l - 1)}{l + 2} \tilde{g} N^{1/2}]^{1/2}.
\end{equation}
We can estimate the strength of the coupling from the 
relation $\tilde{g} = \frac{3 e^{2} b_{z}}{8 \epsilon \hbar \omega}$ 
given in sec.II. In GaAs sample the values of the paraters are 
$b_{z}^{-1} = 58\AA$ and $\epsilon = 12.6$. For $\hbar \omega = 5.4 meV$ 
our estimate of the dimensionless coupling is $\tilde{g} \approx 1.36$. 
For this value of the coupling the variation of excitation energy with 
number of electrons for $l = 1-4$ are shown in figure 2. 

\subsection{Short range interaction}

Now we calculate the sum-rules for the multipole excitations in the case 
of  the short range  interaction. For short-range interaction the density 
profile is given by, 
\begin{equation}
\rho(r) = \frac{1}{\pi l_{0}^{2} \alpha_{s}}(\tilde{\mu} - \frac{1}{2}
x^{2}),
\end{equation}
where, $\alpha_{s} = 1 + \frac{V_{0} m^{*}}{\pi \hbar^{2}}$, and
$l_{0}^{2} =
\frac{\hbar}{m^{*} \omega}$.
The first moment $m^{+}_{1}$ can be written as,
\begin{eqnarray}
m^{+}_{1} & = & m^{*} \frac{\hbar^{2} l^{2}}{m^{*2}} \int d^{2}r r^{2(l -
1)} \rho(r) \nonumber \\
& = & m^{*} \frac{\hbar^{2} l^{2}}{m^{*2}} (\frac{l_{0}^{2(l - 1)}
x_{0}^{2(l + 1)}}{2 \alpha_{s} l(l + 1)}).
\end{eqnarray}
Now let us calculate the third energy weighted moment $m^{+}_{3}$.
Different contributions of this sum-rule are given bellow,
\begin{eqnarray}
m_{3}(T) & = & \frac{2 \hbar^{4} \pi (l - 1)^{2}}{m^{*}} \frac{\hbar^{2}
l^{2}}{m^{*2}} \int d^{2}r r^{2(l - 2)} \rho^{2}(r) \nonumber \\
& = & \frac{\hbar^{4} \pi (l - 1)}{m^{*} \alpha_{s}^{2}l(l + 1)}
\frac{\hbar^{2}
l^{2}}{m^{*2}}l_{0}^{2(l - 3)} x_{0}^{2(l +1)} \\
m_{3}(V) & = & l m^{*} \hbar^{2} \omega^{2} \frac{\hbar^{2} l^{2}}{m^{*2}}
\int d^{2}r r^{2(l - 1)} \rho(r) \nonumber \\
& = & l m^{*} \hbar^{2} \omega^{2} \frac{\hbar^{2} l^{2}}{m^{*2}}
\frac{l_{0}^{2(l - 1)} x_{0}^{2(l + 1)}}{2 \alpha_{s} l(l + 1)} \\
m_{3}(ee) &=& 0.
\end{eqnarray}
From the above expressions collective excitation energy is given by,
\begin{equation}
E_{c}^{2} = \frac{m^{+}_{3}}{m^{+}_{1}} = l \hbar^{2} \omega^{2} + 2(l -
1) \frac{\hbar^{2} \omega^{2}}{\alpha_{s}}.
\end{equation}
Once again for $l=1$, the excitation energy is $E_c=\hbar\omega$ 
consistent with Kohn's theorem. 
For the value of the dimensionless parameter $\alpha_{s} = 3.5$ the 
dispersion relation of the modes $l = 1-4$ are shown in figure 3.

\subsection{Coulomb interaction}
For coulomb interaction we have taken a variational density of the 
following form,
\begin{equation}
\rho(r) = \rho_{0}(x_{0}^{2} - x^{2}),
\end{equation}
where $\rho_{0} = \frac{1}{2 \pi l_{0}^{2} \alpha_{c}}$ and $x = r/l_{0}$.
Here $\alpha_{c}$ is the variational parameter determined by minimizing
the
ground state energy as in Section IIC. This density is similar to that of 
short range interaction. The first moment therefore $m^{+}_{1}$ is given by,
\begin{equation}
m^{+}_{1} = \frac{\hbar^{2} l^{2}}{m^{*}} \frac{\pi \rho_{0} 
r_{0}^{2(l + 1)}}{l_{0}^{2} l (l + 1)}.
\end{equation}

The sum of first two terms of the third moment $m_{3}(T)$ and $m_{3}(V)$
can be written as,
\begin{equation}
m_{3}(T) + m_{3}(V) = \frac{\hbar^{2} l^{2}}{m^{*2}} [ \frac{4 (l - 1) 
\pi^{2} \hbar^{4} \rho_{0}^{2} r_{0}^{2(l + 1)}}{l_{0}^{4} m^{*} l (l + 
1)} + \frac{\pi \hbar^{2} \omega^{2} \rho_{0} m^{*}r_{0}^{2(l + 
1)}}{l_{0}^{2} (l + 1)}].
\end{equation}
We now  calculate $V_{3}(ee)$ for $1/r$ interaction, which can be 
written as,
\begin{eqnarray}
m_{3}(ee) & = & \frac{1}{2} \frac{\hbar^{4} l^{2}}{m^{*2}} [ \int 
d^{2}r \nabla^{2} V_{H} r^{2(l - 1)} \rho(r) + 2(l - 1) \int d^{2}r 
\frac{\partial V_{H}}{\partial r} r^{2l - 3} \rho(r) \nonumber \\ & &
+\int
d^{2}r \int d^{2}r' 
r^{l - 1} \rho'(r) e^{-i l \theta} \frac{1}{|\vec{r} - \vec{r'}|} e^{i l 
\theta'} r'^{l - 1} \rho'(r')].
\end{eqnarray}
where $\rho' = \frac{\partial \rho}{\partial r}$. The hartree potential 
$V_{H}(r)$ is given by,
\begin{eqnarray}
V_{H}(r) & = & \frac{e^{2}}{\epsilon} \int d^{2}r' \frac{1}{|\vec{r} - 
\vec{r'}|} \rho(r) \nonumber \\ 
& = & \frac{e^{2}}{\epsilon l_{0}} \pi \rho_{0} l_{0}^{2} 
(\frac{r_{0}}{l_{0}})^{3} \frac{\Gamma(1/2)}{\Gamma(5/2)} F(1/2, -3/2, 1; 
(\frac{r}{r_{0}})^{2}).
\end{eqnarray}
After performing other integrals and some algebra we arrive at the final 
expression for $m_{3}(ee)$,
\begin{equation}
m_{3}(ee) = \frac{\hbar^{4} l^{2}}{2 m^{*2}} 4 \pi^{2} \rho_{0}^{2}
r_{0}^{2l} 
\frac{e^{2}}{\epsilon l_{0}}(\frac{r_{0}}{l_{0}})^{3} [\frac{4 \Gamma(l + 
1/2)}{ \pi \Gamma(l + 5/2)} - \frac{1}{l + 1} {_{3}}F_{2}(3/2, -1/2, l +
1; l + 2, 2; 1)].
\end{equation}
The collective excitation energy for the multipole modes are given by,
\begin{eqnarray}
E_c^{2}(l) & = & [(l - 1) \frac{2 \hbar^{2} \omega^{2}}{\alpha_{c}} +
l\hbar^{2} 
\omega^{2} \nonumber \\ 
& & + l(l + 1) \frac{e^{2}}{\epsilon l_{0}} \frac{\hbar \omega 
r_{0}}{\alpha_{c} l_{0}} [\frac{4 \Gamma(l + 1/2)}{\pi \Gamma(l + 5/2)} - 
\frac{1}{l + 1} {_{3}}F_{2}(3/2, -1/2,l + 1; l + 2, 2;1)]],\nonumber\\
E_{c}(l) & = & \hbar \omega [2(l - 1)/\alpha_{c} + l + l(l +
1)(\frac{e^{2}}{\epsilon l_{0} \hbar \omega}) \frac{(4
N)^{1/4}}{\alpha_{c}^{3/4}} \times \nonumber\\
& & [\frac{4 \Gamma(l + 1/2)}{\pi \Gamma(l + 5/2)} - \frac{1}{l +
1}~_{3}F_{2}(3/2, -1/2, l + 1;l + 2,2: 1)]]^{1/2},
\end{eqnarray}
where ${_{3}F_{2}(a1,a2,a3;b1,b2;x)}$ denotes the generalised
hypergeometric 
function\cite{grad}.
The validity of Kohn's theorem is easily checked. For the dipole mode $l 
=  1$, we have $_{3}F_{2}(3/2, -1/2, 2; 3, 2;1) = F(3/2, -1/2, 3; 1) = 
\frac{\Gamma(3) \Gamma(2)}{\Gamma(3/2) \Gamma(7/2)}$\cite{grad}.
Substituting this 
into the above equation we obtain the dipole energy, $E_c(1) = \hbar \omega$.

In ref.\cite{serra} the multipole modes of a quantum dot with $\hbar
\omega = 5.6 meV$ containing 56 electrons has been calculated by using
self consistent Hartree-Fock calculation. The multipole excitation
energies for $l = 2 - 4$ are given by, 7.5, 9.0, 10.4 meV respectively.
From our analytical expression we find the corresponding multipole
excitation energies are 7.5, 8.9, 10.12 meV, which are in good agreement
with the self-consistent numerical calculation.
For $\hbar \omega = 5.4 meV$ the multipole modes for $l=1-4$ are plotted 
in the figure 4. Except in the case of dipole mode, the collective excitation 
energy decreases with increasing particle number.

\section{Breathing mode excitation}

In this section we calculate the sum-rules and the excitation energy of 
the breathing mode in a  quantum dot. This excitation is also known as the 
monopole excitation. In the breathing mode excitation average radius 
oscillates around its unperturbed value with a frequency of the excitation 
energy \cite{stringari1}. In a recent experiment on Bose-Einstein condensate 
this phenomena has been obseved very clearly \cite{bec}. 
The monopole excitation operator is given by,
\begin{equation}
F = \sum_{i} (x_{i}^{2} + y_{i}^{2}).
\end{equation}
The first sum-rule $m^{+}_{1}$ can be written as,
\begin{eqnarray}
m_{1} &=& \frac{1}{2} <0|[F^{\dagger}, [\hat{H}, F]]|0> \nonumber \\
&=& \frac{2 \hbar^{2}}{m^{*}} <0|\sum_{i} r_{i}^{2} |0>
\end{eqnarray}
which is nothing but the mean-squared radius in the ground state.

The third moment $m_{3}$ of the strength distribution can be calculated by
using a simple scaling transformation of the ground state 
wave-function\cite{bohigas}. 
Consider the following transformation of the ground state wave function,
\begin{equation}
|\phi_{\eta}> = e^{\eta J} |0>,
\end{equation} 
where $J$ is an anti-harmitian operator $J = [\hat{H}, F]$. Now $J$ can 
be written as, $J = \frac{-2\hbar^{2}}{m}(1 + x\frac{\partial}{\partial 
x} + y\frac{\partial}{\partial y})$. If we take a new set of coordinates 
$x' = log(x)$ and $y' = log(y)$, then the operator $J$ becomes $ \frac{-2 
\hbar^{2}}{m}(1+\frac{\partial}{\partial x'} + \frac{\partial}{\partial 
y'})$. Using these relations, it can be shown that the wave function and the 
coordinates undergo following scaling
transformation,
\begin{equation}
\phi_{\eta}(x, y) = e^{\tilde{\eta}} \phi_{0}(e^{\tilde{\eta}} x,
e^{\tilde{\eta}} y),
\end{equation}
where, $\tilde{\eta} = - \frac{2 \hbar^{2}}{m^{*}} \eta$. The sum-rule $m_{3}$
can be calculated from the scale transformed wave function $\phi_{\eta}$, in the
following way,
\begin{eqnarray}
m_{3} & = & \frac{1}{2} <0|[J^{\dagger}, [\hat{H}, J]]|0> \nonumber \\
& = & \frac{1}{2} \frac{\partial^{2}}{\partial \eta^{2}} <\eta| \hat{H} 
|\eta> |_{\eta = 0}.
\end{eqnarray}
Under this transformation density scales as,
\begin{equation}
\rho_{\eta}(x, y) = e^{2 \tilde{\eta}} \rho(e^{\tilde{\eta}}x,
e^{\tilde{\eta}}y).
\end{equation}
We calculate the sum-rule $m_{3}$ seperately for different terms of the
Hamiltonian. The sum-rule $m_{3}$ can be written as,
\begin{equation}
m_{3} = m_{3}(T) + m_{3}(V) + m_{3}(ee).
\end{equation}
First let us calculate the contribution $m_{3}(T)$ coming from the
kinetic-energy density. Within the Thomas-Fermi approximation we have,
$T[\rho] \approx \int \rho^{2}(r)$. So the kinetic energy of the scaled
density is given by,
\begin{equation}
T[\rho_{\eta}] = e^{2 \tilde{\eta}} T[\rho].
\end{equation}
Using above relations $m_{3}(T)$ can be evaluated in the following way,
\begin{eqnarray}
m_{3}(T) & = & \frac{1}{2} \frac{\partial^{2} T[\rho_{\eta}]}{\partial
\eta^{2}} |_{\eta = 0} \nonumber \\
& = & 2 (\frac{2 \hbar^{2}}{m^{*}})^{2} T[\rho].
\end{eqnarray}
Similarly, for qudratic confinement potential, the potential energy 
scales as $V_{\eta} = e^{-2 \tilde{\eta}} V$, and $m_{3}(V)$ is given by,
\begin{equation}
m_{3}(V) = 2 (\frac{2 \hbar^{2}}{m^{*}})^{2} V.
\end{equation}
Finally we can calculate the term $m_{3}(ee)$ due to the 
electron-electron interaction, in the following way,
\begin{equation}
m_{3}(ee) = \frac{1}{2} \frac{\partial^{2}}{\partial \eta^{2}} 
\frac{1}{2} \int d^{2}r \int d^{2}r' \rho(r) V(\frac{|\vec{r} - 
\vec{r'}|}{\tilde{\alpha}}) \rho(r'),
\end{equation}
where $\tilde{\alpha} = e^{\tilde{\eta}}$.
We now calculate the above quantities and monopole excitation energy for 
different types of interaction.

\subsection{Logarithmic interaction}
Using the density profile of the particles interacting by logarithmic 
potential, the first moment $m_{1}$ for monopole excitation can 
calculated exactly,
\begin{eqnarray}
m_{1} & = & \frac{2 \hbar^{2}}{m^{*}} \int d^{2}r r^{2} \rho(r) \nonumber
\\
& = & \frac{\pi \hbar^{2} \rho_{0} r_{0}^{2}}{m^{*}} [ 1 -
\frac{4}{I_{0}(\tilde{L})}(\frac{I_{1}(\tilde{L})}{\tilde{L}} - \frac{2
I_{2}(\tilde{L})}{\tilde{L}^{2}})],
\end{eqnarray}
where the central density is $\rho_{0} = \frac{1}{\pi l_{0}^{2} 
\tilde{g}}$, and $\tilde{L} = \sqrt{2 \tilde{g}} x_{0}$.
The sum of the first two parts of the third moment $m_{3}$ is given by,
\begin{equation}
m_{3}(T) + m_{3}(V) = 2 (\frac{2 \hbar^{2}}{m^{*}})^{2} [ \frac{\hbar^{2}
\pi}{2 m^{*}}\int d^{2}r \rho^{2} + \frac{1}{2} m^{*} \omega_{0}^{2} \int
d^{2}r r^{2}
\rho(r)]. 
\end{equation}
The third term $m_{3}(ee)$ is given by,
\begin{eqnarray}
m_{3}(ee) & = & \frac{1}{2} \frac{\partial^{2}}{\partial \eta^{2}}
[-\frac{1}{2} \int d^{2}r \int d^{2}r' \rho(r) log(\frac{|\vec{r} -
\vec{r'}|}{\tilde{\alpha} a}) \rho(r)] |_{\eta = 0} \nonumber \\
& = & \frac{1}{2} \frac{\partial^{2}}{\partial \eta^{2}} [
log(\tilde{\alpha}) N^{2}/2 ] = 0,
\end{eqnarray}
where $\tilde{\alpha} = e^{2 \hbar^{2} \eta /m^{*}}$.
The monopole excitation energy is given by,
\begin{equation}
E_{m}^{2} = 2 \hbar^{2} \omega^{2} [1 + \frac{4
f_{m}(\tilde{L})}{\tilde{L}^{2}}],
\end{equation}
where $f_{m}(\tilde{L})$ is given by,
\begin{equation}
f_{m}(\tilde{L}) = \frac{1}{2} \frac{\int_{0}^{1} d^{2}x [1 -
\frac{I_{0}(x\tilde{L})}{I_{0}(\tilde{L})}]^{2}}{\int_{0}^{1} d^{2}x x^{2}
[1 - \frac{I_{0}(x\tilde{L})}{I_{0}(\tilde{L})}]}.
\end{equation}
Using the approximate form for the density profile for $\tilde{L}
>> 1$, the function $f_{m}(\tilde{L})$ is given by,
\begin{equation}
f_{m}(\tilde{L}) = \frac{1 - 2F_{1}(2,3;\tilde{L})e^{-\tilde{L}} +
F_{1}(2,3;2\tilde{L})e^{-2\tilde{L}}}{1 -
F_{1}(4,5;\tilde{L})e^{-\tilde{L}}}.
\end{equation}
The asymptotic $N$ dependence of the breathing mode is given by,
\begin{equation}
E_{m} = \hbar \omega \sqrt{[2 + \frac{4}{\tilde{g}^{2} N}]}.
\end{equation}
In he weak coupling limit$(\tilde{L} << 1)$, the breathing mode excitation energy
is given by,
\begin{equation}
E_{m} = 2 \hbar \omega [1 - \frac{3}{8} \tilde{g} N^{1/2}]^{1/2}.
\end{equation}
For $\tilde{g} = 1.36$, the breathing mode is shown in figure 2. and the 
excitation energy becomes almost constant at the value $\sqrt{2} \hbar 
\omega$. 
This mode has higher excitation energy than the other multipole modes.

\subsection{Short range interaction}
The form of the short range pseudopotential is $ V = V_{0} \delta(r)$, and
the first moment $m_{1}$ for the breathing mode is given by,
\begin{eqnarray}
m_{1} & = & \frac{2 \hbar^{2}}{m^{*}} \int d^{2}r r^{2} \rho(r) \nonumber
\\
& = & \frac{\hbar^{2} l_{0}^{2}}{6 m^{*} \alpha} x_{0}^{6},
\end{eqnarray}
where $\alpha_{s} = 1 + \frac{V_{0} m^{*}}{\pi \hbar^{2}}$, and $x =
r/l_{0}$.
After doing some algebra the third moment $m_{3}$ can be written as,
\begin{eqnarray}
m_{3} & = & m_{3}(T) + m_{3}(V) + m_{3}(ee) \nonumber \\
& = & 2 (\frac{2 \hbar^{2}}{(m^{*})^{2}} [E_{\tau} + E_{p} +
E_{H}]\nonumber\\,
& = & 2 (\frac{2 \hbar^{2}}{m^{*}})^{2}[\frac{\hbar^{2} \pi}{2 m^{*}}
(\frac{x_{0}^{6}}{12 \pi l_{0}^{2} \alpha_{s}^{2}}) + \frac{1}{2} m^{*}
\omega^{2} (\frac{l_{0}^{2} x_{0}^{6}}{12 \alpha_{s}}) + \frac{V_{0}}{2}
(\frac{x_{0}^{6}}{12 \pi l_{0}^{2} \alpha_{s}^{2}})] \nonumber \\
& = & 2 (\frac{2 \hbar^{2}}{m^{*}})^{2} (\frac{m^{*} \omega^{2} l_{0}^{2}
x_{0}^{6}}{12 \alpha_{s}})
\end{eqnarray}
The excitation energy of the breathing mode is given by,
\begin{equation}
E = \sqrt{\frac{m_{3}}{m_{1}}} = 2 \hbar \omega.
\end{equation}
Therefore for the short range interaction the excitation energy is 
independent of the interaction strength.

\section{coulomb interaction}

Now let us consider the case where the electron-electron interaction is
$1/r$ coulomb interaction. Variational density of the $N$ electron system
is taken to be parabolic $\rho = \rho_{0}(x_{0}^{2} - x^{2})$. Where the
central density is $\rho_{0} = \frac{1}{2 \pi l_{0}^{2} \alpha_{c}}$. The
first sum-rule $m_{1}$ is given by,
\begin{eqnarray}
m_{1} & = & (\frac{2 \hbar^{2}}{m^{*}}) \int d^{2}r r^{2} \rho(r)
\nonumber \\
& = & \frac{\pi \hbar^{2} \rho_{0} l_{0}^{4} x_{0}^{6}}{3 m^{*}}.
\end{eqnarray}
Using the scale transformation properties of the density we obtain the
second sum rule $m_{3}$ in the following form,
\begin{eqnarray}
m_{3} & = & \frac{1}{2}(\frac{2 \hbar^{2}}{m^{*}})^{2}[4(T[\rho] + E_{p})
+ E_{H}] \nonumber \\
&=&\frac{1}{2}(\frac{2 \hbar^{2}}{m^{*}})^{2} [\frac{2}{3} \pi \rho_{0} 
l_{0}^{2} x_{0}^{6}(\frac{\pi \hbar^{2} \rho_{0}}{m^{*}} + \frac{1}{2} 
\hbar \omega) + \frac{\Gamma(4) \Gamma(1/2)}{2 (\Gamma(5/2))^{2} 
\Gamma(9/2)} (\frac{e^{2}}{\epsilon l_{0}}) \pi^{2} l_{0}^{4} 
\rho_{0}^{2} x_{0}^{7}].
\end{eqnarray}
From the above expressions of the sum rules, the monopole excitation 
energy($\sqrt{m_{3}/m_{1}}$)can be written as,
\begin{eqnarray}
E^{2} & = & 2 \hbar^{2} \omega^{2} [\frac{1}{\alpha_{c}} + 1 + \frac{3
\Gamma(4) 
\Gamma(1/2)}{4 (\Gamma(5/2))^{2} \Gamma(9/2)}(\frac{e^{2}}{\epsilon l_{0} 
\hbar \omega}) \frac{x_{0}}{\alpha_{c}}],\nonumber \\
E & = & \sqrt{2} \hbar \omega [1/\alpha_{c} + 1 + \frac{128 \sqrt{2}}{105 
\pi} (\frac{e^2}{\epsilon l_{0} \hbar \omega}) 
\frac{N^{1/4}}{\alpha_{c}^{3/4}}]^{1/2}. 
\end{eqnarray}
In Fig.4 the N dependence of the breathing mode is shown for $\hbar 
\omega = 5.4meV$ and it remains constant at the value $1.7 \hbar \omega$.

\section{summary}
In this paper we considered the linear response of quantum dot with finite 
number of electrons under the time dependent perturbation. We have considered 
different types of collective excitations of the dot at zero magnetic 
field. For a suitable excitation operator $F$, the imaginary part of the 
dynamic polarizability (or strength distribution) which is given by,
\begin{equation}
S_{\pm}(E) = \sum_{n} |<0|F_{\pm}|n>|^{2} \delta(E - \hbar \omega_{n})
\end{equation}
becomes a highly peaked function near the collective excitation energy. 
The collective excitation energy is well estimated by the energy weighted 
sum-rule approach, and is given by, $E_{coll} = \sqrt{\frac{m_{3}}{m_{1}}}$.
We have disscussed in general both multipole excitations(also known as edge modes) 
and breathing mode excitations in a quantum dot. To calculate the collective 
excitation energies we have taken three different types of electron-electron 
interaction, which are, the logarithmic interaction, the short range 
pseudopotential and the coulomb interaction. Within the Thomas-Fermi 
approximation 
the ground state density has been calculated. This is a good approximation 
for large 
number of particles. 
For multipole excitations the angular momentum of the system changes, and
the the excitation operator is defined as $z^{l}$, which is the basis for
the expansion of the angular part of any density fluctuation in
2-dimension. In quantum dot with large number of electrons and logarithmic
two body interaction, multipole excitation energies form a band like
structure around the energy $\hbar \omega$, and the energy spacing goes
as, $\sim 1/r_{0}^{2}$. For short range interaction excitation energies
are $ E = \hbar \omega \sqrt{l + 2(l - 1)/\alpha_{s}}$, which is
independent
of the number of electrons $N$, but depends on the
interaction parameter $\alpha_{s}$. For coulomb interaction excitation
energies take the following form,
\begin{eqnarray}
E^{2} & = & \hbar^{2} \omega^{2} [l + (l - 1)/\alpha_{c} \nonumber
\\ & & + l(l + 1)
(\frac{e^{2}}{\epsilon l_{0} \hbar \omega}) \frac{r_{0}}{\alpha_{c} l_{0}}
[\frac{4 \Gamma(l + 1/2)}{\pi \Gamma(l + 5/2)} - \frac{1}{(l + 1)}
{_{3}}F_{2}(3/2,-1/2,l + 1; l + 2, 2; 1)].
\end{eqnarray}
For dipole mode all results are consistent with Kohn's theorem, and the dipole
excitation energy is $\hbar \omega$.

For the breathing mode excitation, average value of the collective coordinate
$r^{2}$ oscillates around its equilibrium value. For logarithmic
interaction breathing mode excitation energy becomes $\sqrt{2} \hbar
\omega$ in large $N$ limit. For short range interaction excitation energy
is $2 \hbar \omega$, which same as the excitation energy of the
non-interacting system. For coulomb interaction the breathing mode remains
almost constant at the value $1.7\hbar\omega$. As shown in the 
Appendices, the low lying excitations are well described by the T-F 
approximation for the ground state density in the large $N$ limit. The 
corrections may also be calculated in order to correct for the 
inaccqurecies that may arise for higher excitations.

Apart from the single particle excitations, collective excitations in
large quantum dots and their dependence on the number of particles and
external magnetic field has become very interesting topic. Detailed study
of
the collective excitations can give information about the
electron-electron correlation in finite size quantum dot.

I would like to thank M. V. N. Murthy for his helpful comments and critical 
reading of the manuscript. I would also like to thank R. K. Bhaduri for
his helpful comments.

\appendix
\section{}

In this appendix we describe the corrections of the collective frequencies
due to the edge corrections of the T-F density for a simple case. Since
for the logarithmic two-body interaction, the T-F density can be written
in a closed form, we take this example. The density distribution for
logarithmic interaction is given by,
\begin{equation}
\rho(r) = \frac{1}{\pi l_{0}^{2} \tilde{g}}[1 -
\frac{I_{0}(\sqrt{2\tilde{g}} x)}{I_{0}(\sqrt{2\tilde{g}} x_{0})}],
\end{equation}
where, $x = r/l_{0}$, and $r_{0}$ is the turning point.
When $\tilde{L} = \sqrt{2\tilde{g}} x_{0} >> 1$, then the density becomes
almost flat. In this limit we can approximate the T-F density by a
"Fermi-distribution" like density, which has a diffusive tail also and this
density is given by,
\begin{equation}
\rho_{a}(r) = \frac{\rho_{0}}{1 + exp(\frac{r - R}{a})},
\end{equation}
where, $\rho_{0} = \frac{1}{\pi l_{0}^{2} \tilde{g}}$, $R \approx r_{0}$, and $a \approx 
l_{0}/\sqrt{2\tilde{g}}$. The multipole excitation energies are given by,
\begin{equation}
E_{c}^{2} = \hbar^{2} \omega_{0}^{2} + \frac{3\hbar^{4}\pi (l - 1)^{2} Q_{2}(l)}{m^{*2} 
Q_{1}(l)},
\end{equation}
where, $Q_{2}(l) = \int r^{2l - 3} \rho_{a}(r)^{2} dr$, and $Q_{1}(l) = \int 
r^{2l - 1} \rho_{a}(r) dr$.
Now we can expand $Q_{1}(l)$ and $Q_{2}(l)$ in a series of $a/R$, which is known as 
leptodermous expansion. Upto the leading order correction $Q_{1}$ and
$Q_{2}$ are given by \cite{treiner},
\begin{eqnarray}
Q_{1}(l) & = & \frac{\rho_{0} R^{2l}}{2l}[1 + 2l(2l - 1) 
\frac{\pi^{2}}{6}(\frac{a}{R})^{2}], \\ 
Q_{2}(l) & = & \frac{\rho_{0}^{2} R^{2(l - 1)}}{2(l - 1)}[1 - 2(l - 
1)(\frac{a}{R})],
\end{eqnarray}
if we set $ a = 0$ we recover the results given in section IV.
Using the above results, upto the leading order in $a/R$, collective 
frequencies are given by, 
\begin{equation}
E_{c} = \hbar \omega_{0}[\sqrt{1 + \frac{3l(l-1)}{\tilde{g}^{2} N}} + 
\frac{6l(l-1)^{2}}{(2 \tilde{g}^{2} N)^{3/2}}].
\end{equation}
So the leading order correction goes as, $\sim (l/\sqrt{N})^{3}$, which 
is negligible for low lying excitations in a large system.

%\appendix
\section{}

In this appendix we estimate the correction of two important moments 
of the density, due to the diffusive edge of the density. These moments 
essentially controll the excitation energies we are interested in. To 
include 
the asymptotic tail of the density, we write the total density in the 
following way,
\begin{equation}
\rho(r) = \rho_{TF}(r) + \delta \rho(r),
\end{equation}
where $\rho_{TF}(r)$ is the Thomas-Fermi density, and $\delta \rho$ 
is the edge correction. Since the quantum mechanical density of the
electrons in the harmonic oscillator potential asymptotically decays as
$\sim e^{-r^{2}/l_{0}^{2}}$, we choose the following form of the edge
correction,
\begin{equation}
\delta \rho(r) = \delta \rho_{0} e^{(r_{0} - a)^{2} /l_{0}^{2}} 
e^{-r^{2} / l_{0}^{2}}~~~~~~~~for~ r>r_{0} - a,
\end{equation}
where, $a$ is the size of the region near the turning point $r_{0}$ where
the T-F density deviates from the actual quantum mechanical density. In
the large $N$ limit $a << r_{0}$.At $r = r_{0} - a$ we match two
densities, $\rho_{T-F}(r_{0} - a) = \delta \rho_{0}$. Now we consider the
following moments, \begin{eqnarray}
I_{1}(n)& = & \int r^{n} \rho(r) d^{2}r  \\
I_{2}(n)& = & \int r^{n} \rho^{2}(r) d^{2}r.
\end{eqnarray}
The corrections of the moments due the asymptotic tail of the density
distribution may be written as,
\begin{eqnarray}
\delta I_{1}(n) & \approx & \delta \rho_{0} e^{r_{0}^{2}/l_{0}^{2}}
\int_{r_{0}}^{\infty} d^{2}r r^{n} e^{-r^{2}/l_{0}^{2}} \\
\delta I_{2}(n) & \approx & (\delta \rho_{0})^{2} e^{2
r_{0}^{2}/l_{0}^{2}}
\int_{r_{0}}^{\infty} d^{2}r r^{n} e^{-2 r_{0}^{2}/l_{0}^{2}}.
\end{eqnarray}
By using parabolic T-F density, we can calculate $(I_{1}(n))_{TF}$ and 
$(I_{2}(n))_{TF}$. The leading order fractional change of the moments 
due to the edge correction can be written as,
\begin{eqnarray}
|\frac{\delta I_{1}(n)}{(I_{1}(n))_{TF}} | & \approx & \frac{(n + 2)(n + 
4)}{2} (\frac{a}{l_{0}})~ \frac{1}{\tilde{r}_{0}^{3}} \nonumber \\
& = & \frac{(n + 2)(n + 4)}{2} (\frac{a}{l_{0}})~ \frac{1}{(4 N)^{3/4}},
\end{eqnarray}
where $\tilde{r_{0}} = r_{0}/l_{0}$, for noninteracting electrons
$\tilde{r_{0}} = (4 N)^{1/4}$. The correction of $I_{2}(n)$ is given by,
\begin{eqnarray}
|\frac{\delta I_{2}(n)}{(I_{2}(n))_{TF}}| & \approx & (n/2 + 1)(n/2 +
2)(n/2 +
3) (\frac{a}{l_{0}})^{2}~\frac{1}{\tilde{r}_{0}^{3}}~[
(4\frac{a}{l_{0}}) +
\frac{1}{\tilde{r}_{0}}] \nonumber \\
& \approx & (n/2)^{3} (\frac{a}{l_{0}})^{2}~ \frac{1}{(4
N)^{3/4}}~[(4\frac{a}{l_{0}}) + \frac{1}{\tilde{r}_{0}}].
\end{eqnarray}
Therefore the errors are at most of the order of $n^{3}/N^{3/4}$. We may
now estimate $a/l_{0}$ in the following way. The wave length $\lambda$ of
the particles near the fermi surface is of the order of $\sim
\frac{\hbar}{P_{F}(r)}$. The local fermi momentum is given by,
$P_{F}(r) = \sqrt{2m (\mu - V(r))}$. The semiclassical approximation
breaks down when $|\frac{\partial \lambda(r)}{\partial r}| \sim 1$. From
this condition we obtain $a/l_{0} \sim \frac{1}{2 \tilde{r}_{0}^{1/3}} =
1/(2 (4 N)^{1/12})$. This estimate shows that the error can be reduced to
even smaller value in the large $N$ limit.

A more direct way of estimating corrections to the T-F
limit is to compare the results with those obtained using the quantum
mechanical density. This is rather hard in the case of interacting system,
but easier in the non-interacting limit. The exact quantum mechanical
density can be written as a sum over the squares of various orbital
wavefunctions, $\rho(r) = \sum_{n,l} |\psi_{n, l}(r)|^2$, where the sum is
restricted so that the number of orbitals included conserve number of
particles $N$.
We calculated the excitation frequencies for two values
of $N$, by using quantum mechanical density, and compared them with
frequencies obtained from T-F density. For $N=12$ and $l=4$ excitation
frequency T-F model deviates by only $10$ percent, for the lower
frequencies
error is much less. For $N=20$ and $l = 4$ collective mode the error due
to semiclassical approximation reduces to $6$ percent. This simple
analytical
calculation supports our large $N$ asymptotic analysis of the error due to
the use of semiclassical density. If we consider the low-lying collective
excitations in a large quantum system, then our calculation indicates that
the error in the collective modes decreases asymptotically with the
increasing number of particles.

\newpage
\begin{figure}
\caption{Comparision of the density profile of electrons with logarithmic 
interaction with the parapolic density profile for two different values of
$\tilde{L}$. Solid line shows the density of the electrons for $\tilde{L} 
= 1$, dashed line shows the density for $\tilde{L} = 20$, and dot-dashed 
line shows the parabolic density.}

%\begin{figure}
\caption{Excitation energies($\hbar \omega_{c}$) as a function of the number of 
particles (N), for logarithmic interaction with $\tilde{g} = 1.36$. Solid line 
shows the exact result and the dashed line shows asymptotic value of the excitation 
energy. $l$ values indicate the corresponding multipole modes and $n=1$
level indicate 
the breathing mode.}

%\begin{figure}
\caption{Excitation energies($\hbar \omega_{c}$) as a function of the number of 
particles (N), for short range interaction with $\alpha_{s} = 3.5$.
Multipole modes are denoted by corresponding $l$ values and $n=1$ level indicate
the
breathing mode} 

%\begin{figure}
\caption{Excitation energies($\hbar \omega_{c}$) as a function of the 
number of particles (N), for coulomb interaction with 
$\frac{e^{2}}{\epsilon l_{0} \hbar \omega} = 1.45$. Multipole modes are 
denoted by 
corresponding $l$ values and $n=1$ level indicate the breathing mode.}

\end{figure}


\bigskip

\begin{references}

\bibitem{lipparini}
E. Lipparini and S. Stringari, phys. Rep. {\bf 175}, 103 (1989).

\bibitem{stringari1}
S. Stringari et al, Nuclear Phys. A, {\bf 309}, 177 (1978).

\bibitem{brack}
M. Brack, Rev. Mod. Phys,{\bf 65}, 677 (1993).

\bibitem{deheer}
W. de Heer, Rev. Mod. Phys,{\bf 65}, 611 (1993).

\bibitem{merket}
Ch. Sikorski and U. Merkt, Phys. Rev. Lett. {\bf 62}, 2164 (1989).

\bibitem{demel}
T. Demel, D. Heitmann, P. Grambow, and K. Ploog, Phys. Rev. Lett. {\bf
64}, 788 (1990).

\bibitem{schuller}
C. Schuller, K. Keller, G. Biese, E. Ulrichs, L. Rolf, C.Steinebach, and 
D. Heitmann, Phys. Rev. Lett. {\bf 80}, 2673 (1998). 

\bibitem{tapash}                                                                
P. A. Maksym and T. Chakraborty, Phys. Rev. Lett. {\bf 65}, 108 (1990).

\bibitem{serra}
E. Lipparini, N. Barberan, M. Barranco, M. Pi, and L. Serra et al, Phys. 
Rev. B, {\bf 56}, 12375 (1997); 
A. Emperador, M. Barranco, E. Lipparini, M. Pi, and L. Serra cond-mat/9902357.

\bibitem{stringari2}
S. Giovanazzi, L. Pitaevskii and S. Stringari, Phys. Rev. Lett. {\bf 72}, 
3230 (1994).   

\bibitem{mcdonald}
A. H. MacDonald, J. Phys. C, {\bf 18}, 1003 (1985).
  
\bibitem{fetter}
D. B. Mast, A. J. Dahm, and A. L. Fetter, Phys. Rev. Lett. {\bf 54}, 1706
(1985).

\bibitem{sikin}
V. Shikin, S. Nazin, D. Heitmann, and T. Demel, Phys. Rev. B, {\bf 43}, 
11903(1991).

\bibitem{peeters}
B. Partoeus, A. Matulis and F. M. Peeters, cond-mat 9712066

\bibitem{matthias}
Matthias Brack and Rajat. K. Bhaduri, Semiclassical Physics,
Addison-Wesley Publishing  Company.

\bibitem{pino}
Ramiro Pino, Phys. Rev. B, {\bf 58}, 4644 (1998).

\bibitem{sinha}
S. Sinha, M. Sc thesis, (unpublished).

\bibitem{zhang}
F. C. Zhang and S. DasSarma, Phys. Rev. B, {\bf 33}, 2903 (1986).

\bibitem{grad}
I. S. Gradshteyn and I. M. Ryzhik, Tables of integrals, Series and Products 
(Academic Press, New York, 1980).

\bibitem{trugman}
S. A. Trugman and S. Kivelson, Phys. Rev. B, {\bf 31}, 5280 (1985).

\bibitem{murthy}                                                                
R. K. Bhaduri, M. V. N. Murthy, and M. K. Srivastava, Phys. Rev. Lett.          
{\bf 76}, 165 (1996).

\bibitem{fogler}
M. M. Fogler, E. I. Levin, and B. I. Shklovskii, Phys. Rev. B, {\bf 49},
13767 (1994).

\bibitem{kohn}
W. Kohn, Phys. Rev. {\bf 123}, 1242 (1961).

\bibitem{bec}
M. O. Mews et al, Phys. Rev. Lett, {\bf 77}, 988 (1996).

\bibitem{bohigas}
O. Bohigas, A. M. Lane, and J. Martorell, Phys. Rep. {\bf 51}, 267 (1979).

\bibitem{treiner}
J. Treiner and H. Krivine, Annals of Physics, {\bf 170}, 406 (1986).

\end{references}
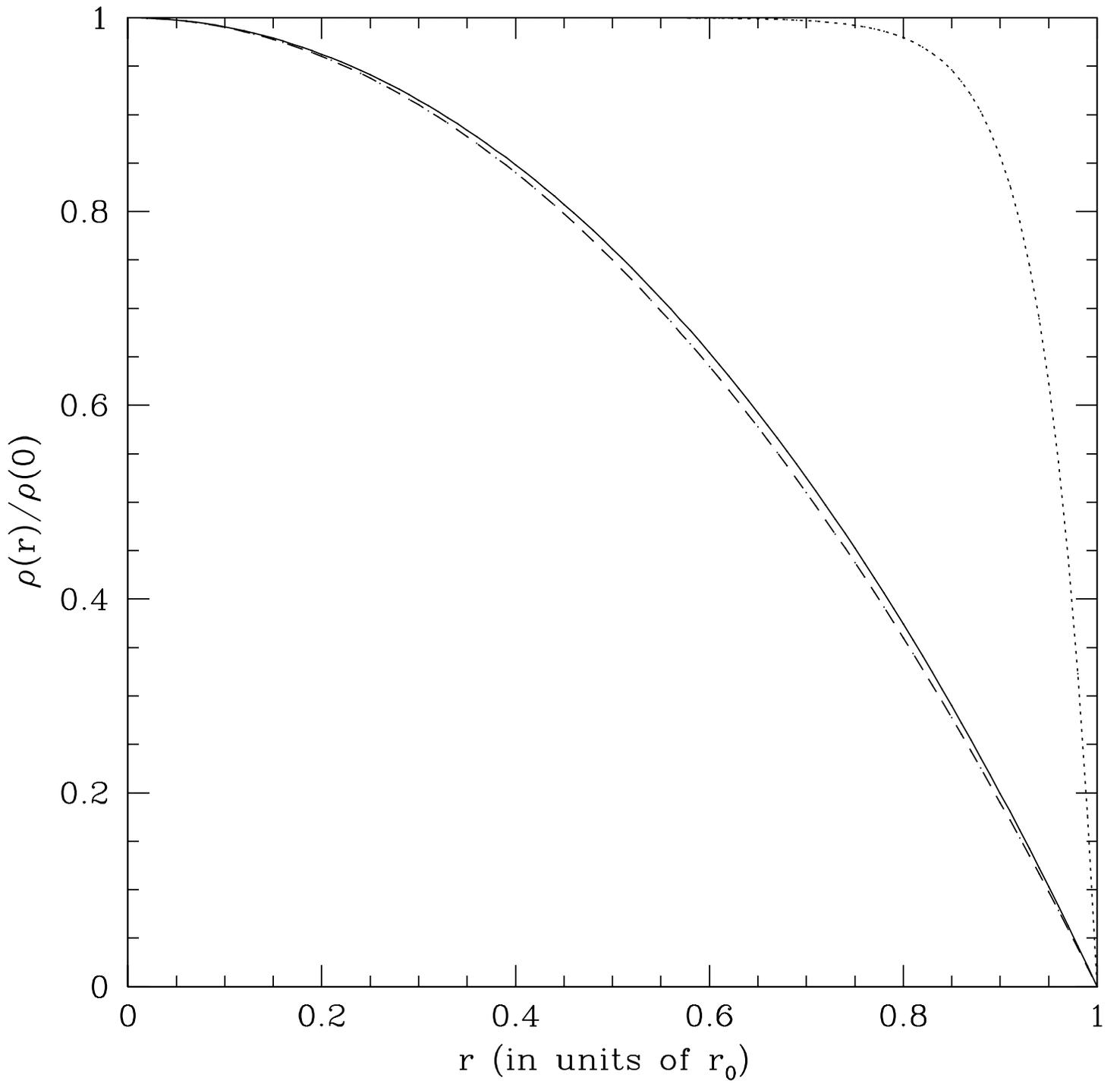
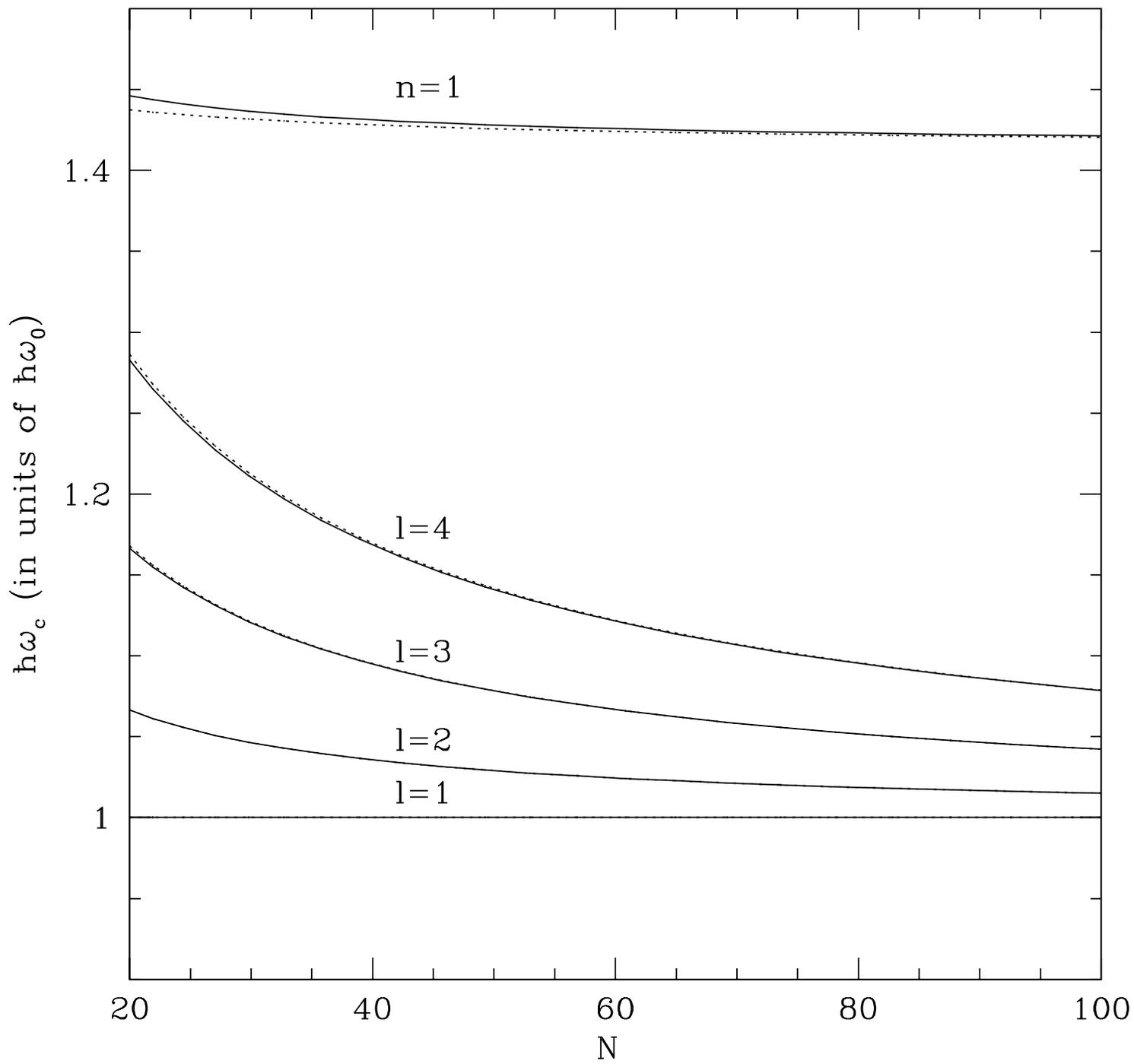
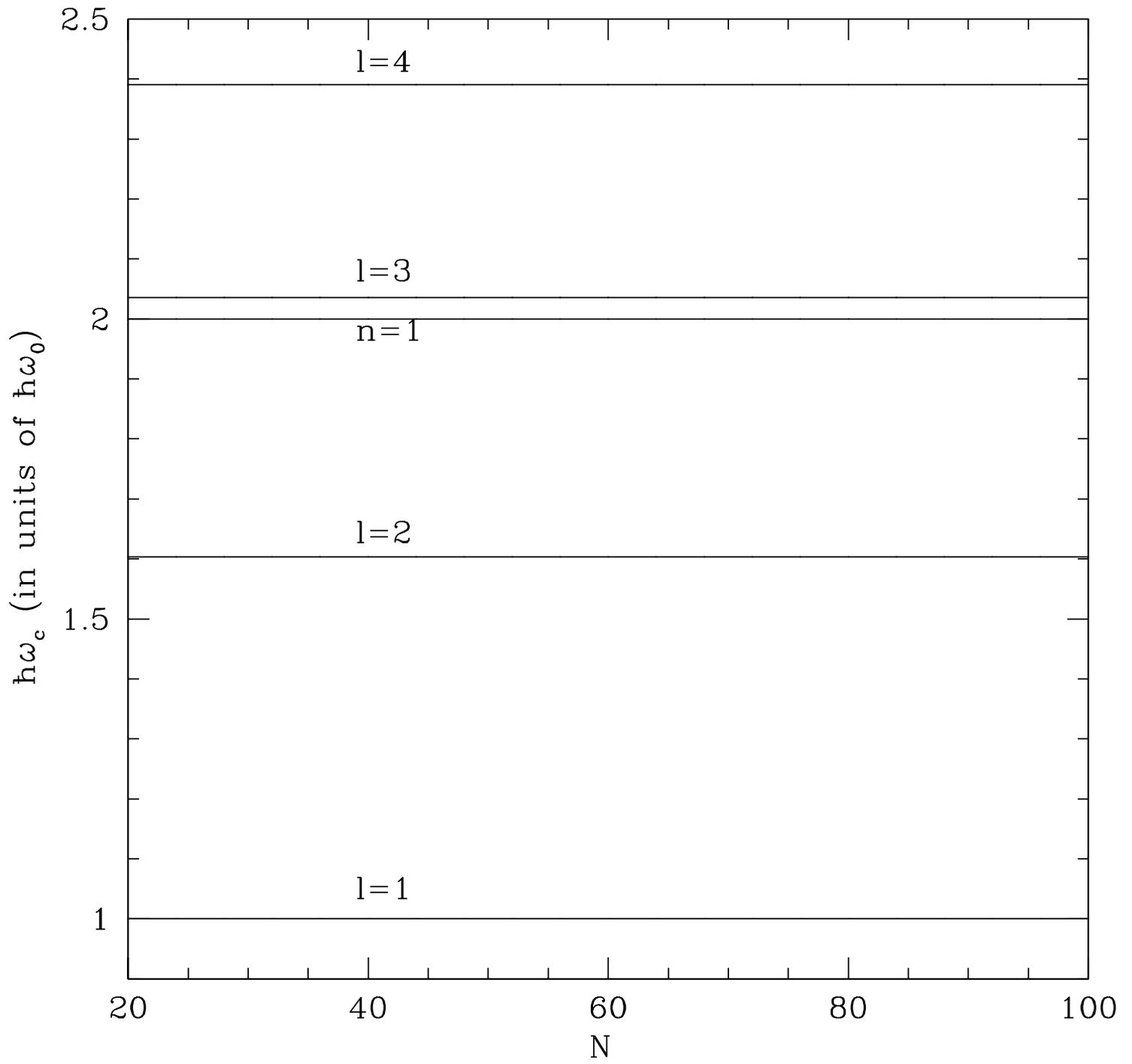
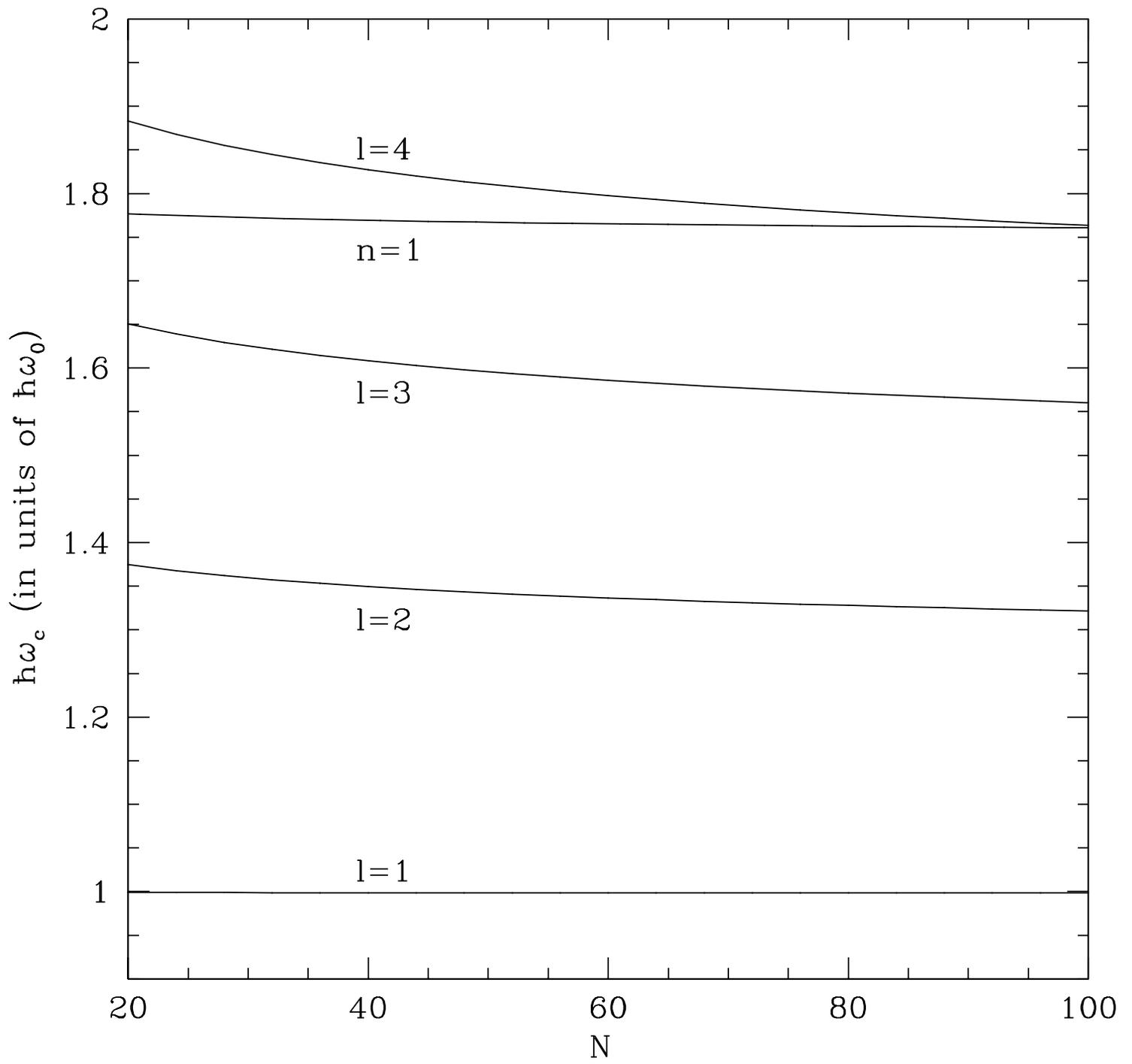
\end{document}